\documentclass[12pt,aps,preprint,nofootinbib]{revtex4}
\usepackage{graphicx}
\usepackage{cancel}
\usepackage{amssymb}
\usepackage{amsmath}
\usepackage{bm}
\usepackage{euscript,amsmath}
\def\lslash{\rlap{\hspace{0.02cm}/} {l}}

\def\beq{\begin{equation}}
\def\eeq{\end{equation}}
\def\bea{\begin{eqnarray}}
\def\eea{\end{eqnarray}}
\def\to{\rightarrow}
\def\lsim{\mathrel{\rlap{\lower4pt\hbox{\hskip1pt$\sim$}}
    \raise1pt\hbox{$<$}}}         
\def\gsim{\mathrel{\rlap{\lower4pt\hbox{\hskip1pt$\sim$}}
    \raise1pt\hbox{$>$}}}         

\begin{document}
\preprint{IPMU-09-0103,ZIMP-09-04}
\title{Invisible Higgs decay with $B\to K\nu\bar{\nu}$ constraint}
\author{C.S. Kim~$^{\bf a,b}$}
\email{cskim@yonsei.ac.kr}
\author{Seong Chan Park~$^{\bf b}$}
\email{seongchan.park@ipmu.jp}
\author{Kai Wang~$^{\bf b}$}
\email{kai.wang@ipmu.jp}
\author{Guohuai Zhu~$^{\bf c}$}
\email{zhugh@zju.edu.cn}
\affiliation{
$^{\bf a}$ Department of Physics, Yonsei University, Seoul 120-479, KOREA\\
$^{\bf b}$ Institute for the Physics and Mathematics of the Universe, University of Tokyo, Kashiwa, Chiba 277-8568, JAPAN\\
$^{\bf c}$ Zhejiang Institute of Modern Physics and Department of Physics, Zhejiang University, Hangzhou, Zhejiang 310027, CHINA
}

\begin{abstract}
If the Higgs boson were the only particle within the LHC accessible range, precision measurement of the Higgs's properties would play a unique role
in studying electroweak symmetry breaking as well as possible new physics. We try to use low energy experiments such as rare $B$ decay to constrain
a challenging decay mode of Higgs, in which a Higgs decays to a pair of light ($\approx 1 \sim 2$ GeV) SM singlet $S$ and becomes invisible. By using the current experimental bound of rare decay $B\to K\nu\bar{\nu}$ and computing the contribution of $B\to K SS$ to (the) $B\to K+\cancel{E}$, we obtain an upper bound on the Higgs coupling to such light singlet. It is interesting that the partial width of the invisible decay mode $h\to SS$ by taking the upper bound value of coupling is at a comparable level with $h\to WW/ZZ$ or $WW^{(*)}$ decay modes,making the Higgs identifiable but with a different predicted decay BR from [the] standard model Higgs decay. It will then have an impact on precision measurement of the Higgs's properties. We also study the implication for cosmology from such a light singlet and propose a solution to the potential problem.
\end{abstract}
\maketitle

\section{Introduction}

Searching the Higgs boson, the last missing piece in the Standard Model (SM)
of particle physics, is one of the essential goals of the CERN Large Hadron Collider (LHC).
The Minimal Higgs boson model is the simplest solution to electroweak symmetry
breaking and also the most economic one to be consistent with existing precision
measurements. However, theoretical considerations suggest
that the Minimal Higgs boson model may not be complete.
Being a fundamental scalar, Higgs boson receives quantum corrections of quadratic
divergence. To solve this, there have been many theoretical proposals
which predict various new physics at $\cal O$(TeV). Direct evidences of new resonances
at the LHC can determine what the new physics model is.
However, if the Higgs boson were the only particle at
the LHC accessible range, we will have to
rely on precision measurements. The precision
measurement of Higgs boson properties can play an
important role to confirm electroweak symmetry breaking mechanism \cite{Duhrssen:2004cv}
and test new physics \cite{Low:2009di}.
For instance, measurement on
top quark Yukawa coupling $y_t$ is crucial to probe
the origin of fermion mass generation while
$gg\to h$ production due to the top quark loop directly depends on the coupling $y_t$.
On the other hand, the contribution from new physics may also change the
$gg\to h$ production rate significantly.
One interesting scenario will be
that at the LHC one does discover the conventional
Higgs search channels, confirm it is the Higgs and
measure its mass but the observed event number is
much smaller than what we expect for the SM Higgs
of that mass.

However, a new decay mode of Higgs boson
that cannot be easily identified will lead to the
same consequence when the new decay width
is comparable with the conventional SM Higgs
width at the same mass \cite{neutrino}.
For instance, if Higgs decay has an invisible mode,
it is impossible to fully reconstruct such resonance
and is very challenging
to identify at the hadron colliders \cite{invisible}.

In this paper, we want to consider the invisible decay of Higgs to
a pair of hidden sector scalar ($S$) particles in the minimal extension
of the SM \cite{Silveira:1985rk, Davoudiasl:2004be,Bird:2004ts,SungCheon:2007nw}.
As the scalar particle is a singlet of the SM interactions it can
only directly couple to the Higgs by the interaction Lagrangian
\beq
{\lambda \over 2v_0} H^\dagger H S^2 \equiv {\hat{\lambda} \over 2} H^\dagger H S^2,
\eeq
where $\lambda$ is a dimension one coupling constant and $v_0$ the vacuum
expectation value (vev) of the Higgs boson.
It is a challenge to identify such invisible Higgs at collider
experiments and
obtain any bound on invisible Higgs. The only controlled experiments at
this moment that
can put constraints on such decay mode are through low energy
processes such as rare $B$ or $K$ decays. In these processes the Higgs is virtual, not interacting
directly to $B$ or $K$, but to top quark and $S$. Therefore, the only difference is CKM factor, for $K$
it is about $10^{-5}$ smaller than $B$, so we would need more than $10^{10}$ $K$'s.
Therefore, we just focus on rare $B$ decays in this work.

In Table~\ref{table:1}, we show the theoretical estimates of branching ratios (BRs)
within the SM \cite{Jeon:2006nq,Altmannshofer:2009ma,Kamenik:2009kc,Bartsch:2009qp}
and their current experimental bounds at $B$ factories \cite{:2007zk,Aubert:2004ws,:2008fr}
for the decays $B \to M \nu \bar \nu$.
The errors of the SM estimates in Table~\ref{table:1} are
mainly due to the hadronic transition form factors and the
CKM matrix elements, since those decay channels are  among
the cleanest SM processes due to only
involving electroweak penguin diagrams \cite{Buchalla:1993bv}, except for $B \to \pi \nu\bar\nu$  \cite{Kamenik:2009kc}.
Please note that by taking the ratios such as $\text{Br}(B \to \pi \nu \bar\nu)/\text{Br}(B \to \pi l \nu)$,
$\text{Br}(B \to K^* \nu \bar\nu)/\text{Br}(B \to \rho l \nu)$, we can reduce considerably
the uncertainties related to the hadronic form factors \cite{Aliev:1997se}. For $B \to K \nu \bar{\nu}$,
similarly one may consider the ratio $\text{Br}(B \to K \nu \bar\nu)/\text{Br}(B \to K \ell^+ \ell^-)$
where the uncertainties from the hadronic form factors are canceled to a large extent \cite{Bartsch:2009qp}.

\begin{table}
\caption{Expected BRs in the SM and  experimental bounds (90\% C.L.)
in units of $10^{-6}$. The SM values for $K,\pi,K^*$ include the long distance contributions through intermediate
on-shell $\tau$, which can be dominant for $\pi$ case \cite{Kamenik:2009kc}.
}
\smallskip
\begin{tabular}{|c|c|c|}
\hline
~~mode~~ & ~~BRs in the SM~\cite{Jeon:2006nq,Altmannshofer:2009ma,Kamenik:2009kc,Bartsch:2009qp}~ & ~~Experimental bounds~~ \\
\hline
~~$B\to K\nu\bar{\nu}$~~       & $5.1 \pm 0.8$
                               & $<\phantom{1}14$ \cite{:2007zk} \\
~~$B\to \pi\nu\bar{\nu}$~~     & $9.7 \pm 2.1$
                               & $<100 $ \cite{Aubert:2004ws} \\
~~$B\to K^\ast \nu\bar{\nu}$~~ & $8.4 \pm 1.4$ & $<80 $ \cite{:2008fr} \\
~~$B\to \rho \nu\bar{\nu}$~~   & $0.49^{+0.61}_{-0.38}$  & $<150 $ \cite{:2007zk} \\
   \hline
   \end{tabular}
   \label{table:1}
\end{table}

Here we will focus on $B^+\to K^+\nu\bar{\nu}$ decay as its experimental upper bound is
closest to the SM prediction as shown in Table~\ref{table:1}.
Using the SM expectation value
\beq
\text{Br}_\text{SM}(B^+ \to K^+ \nu \bar{\nu}) = 5.1\pm 0.8 \times 10^{-6},
\eeq
and the current upper bound from BELLE \cite{:2007zk}
on this final state as
\begin{equation}
 \text{Br}(B \to K + \cancel{E}) < 14 \times 10^{-6}~,
\end{equation}
we can derive the corresponding constraint on
Higgs invisible decay width.

To be kinematically allowed in $B\to K SS$,
the singlet scalar cannot be heavier than $1-2$ GeV.
Therefore, the scalar can be easily thermalized through the Higgs interactions in the early
universe. We first discuss its cosmological bound in next section.
The third section is the discussion on $B$ decay. After taking into all
the constraints, we discuss its implication in Higgs in the Section IV and
finally the conclusion in Section V.

\section{Cosmological bound and decay of a hidden sector scalar}

If we assume the renormalizability of the theory and allow the mass term  quartic self-interaction term
and the quartic interaction term with the Higgs, the Lagrangian of the scalar sector  is written as
\begin{eqnarray}
{\cal L}_{\rm scalar} = \frac{1}{2}(\partial S)^2 -\frac{1}{2}m_S^2 S^2 -\frac{\lambda_S}{4!}S^4 -\frac{\hat{\lambda}}{2}S^2 H^\dagger H.
\end{eqnarray}
The Lagrangian respects the $Z_2$ symmetry ($S\to -S$) thus $S$ is a stable particle.
Indeed this scalar particle can be a good candidate of dark matter.
The scalar particle could be in thermal equilibrium with the SM sector through interaction
with Higgs boson in early universe and finally its relic still may survive
in the current universe in the form of dark matter \cite{Silveira:1985rk, Burgess:2000yq}.
The relic density is determined by annihilation cross section of the scalar particle to
the SM particles as \cite{Bertone:2004pz}
\begin{eqnarray}
\Omega_S h^2 \simeq \frac{0.1 {\rm pb}}{\langle \sigma_S v_{rel}\rangle},
\end{eqnarray}
where $\sigma_S$ is the annihilation cross section of $S$ to the standard model particles
through $s$-channel Higgs exchange diagrams and $v$ is relative velocity between annihilating $S$s.
Since we are mainly interested in GeV scale particle, available channels are mainly
to light leptons ($e,\mu, (\tau)$) and quarks ($u, d, s, c (,b)$) and the cross section is obtained as
\begin{eqnarray}
\langle \sigma_S v_{rel}\rangle =\frac{\hat{\lambda}^2 m_S^2}{ \pi m_h^4} \Phi(m_S).
\end{eqnarray}
The precise value of $\Phi(m_S) \simeq  \sum_f x_f^2 (1-x_f^2)^{3/2}$ where $x_f=m_f/m_S$ depends
on the actual mass of scalar particle and the kinematically allowed channels.
We found a stringent constraints on the annihilation cross section considering
the WMAP data $\Omega_c h^2 = 0.1131\pm 0.0034$ \cite{wmap5yr} as
\begin{eqnarray}
\frac{\hat{\lambda}^2 m_S^2}{\pi m_h^4} &&\gsim \frac{0.1 {\rm pb}}{\Omega h^2|_{\rm WMAP5yr}}\nonumber \\
\Rightarrow\,\, \hat{\lambda} \gsim 3.5 &&\times \left(\frac{{\rm 1 \,GeV}}{m_S}\right)\times
\left(\frac{m_h}{150 \,{\rm GeV}}\right)^2.
\end{eqnarray}
If $m_h\simeq 150 (115)$ GeV and $m_S\simeq 1$ GeV we get $\hat{\lambda} \gsim 3.5 (1.2)$, respectively,
which is within the strong coupling regime where the perturbative description of the model is not available.

\begin{figure}[t]
\includegraphics[scale=1.0,width=10cm]{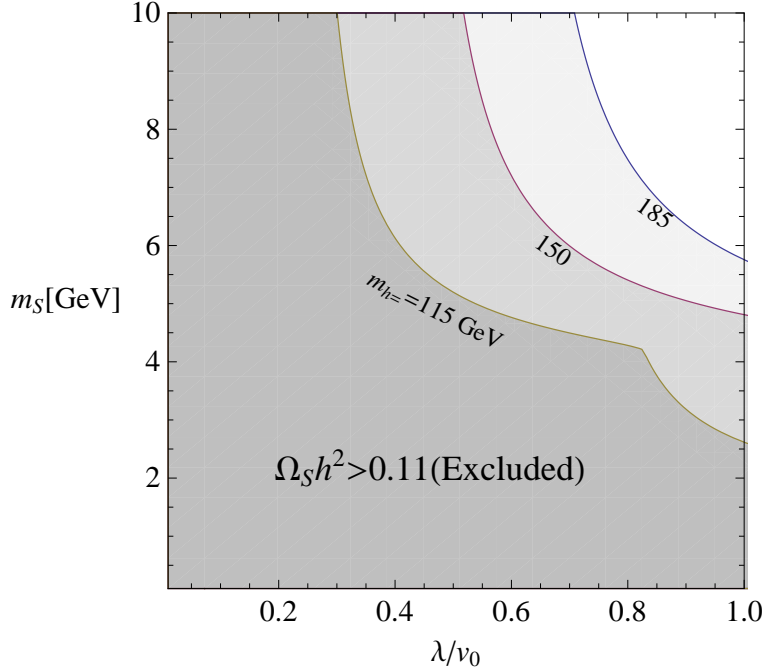}
\caption{Cosmological constraints for a stable $S$ from the relic abundance. Allowed parameter space in $(\hat{\lambda}=\lambda/v_0, m_S)$ plane with $m_h=115, 150$ and $185$ GeV, respectively.  }
\label{fig:relic}
\end{figure}

 In Fig. \ref{fig:relic}, we presented the allowed parameter space in $(\hat{\lambda}=\lambda/v_0, m_S)$
 plane by the 5 year WMAP data on CDM component with various values of Higgs mass $(115,150,185)$ GeV
 taking threshold effects into account. Basically a GeV scale mass range, only in which range $B\to K SS$ is allowed,
 is not compatible with the cosmological observations
 \footnote{In Ref. \cite{Bento:2001yk}, a scalar field in the mass range
of $1$ GeV has been considered and the authors reached the same
conclusion with ours: a large coupling constant is required in
order to avoid overabundance. However, this large coupling
constant is ruled out by the $B\to K\nu\bar{\nu}$ data, as we consider in Sec. III.}.
On the other hand, if the scalar is heavier ($m_S>2$ GeV)
 even though the scalar cannot contribute to the $B$-decays but can be a successful dark matter candidate,
 if the $\lambda$ coupling is properly chosen.


However, we can easily avoid this cosmological constraint provided that the singlet actually decays
into light particles since only (absolutely) stable particles can significantly contribute
to the dark matter density of the current universe.  As the longevity of the scalar particle is inherited
by the $Z_2$ symmetry, a mechanism of breaking $Z_2$ symmetry leads to a natural way out.
Indeed there is a very promising source of the symmetry breaking.  Quantum gravity effect
actually allows higher order operators and some of them might break global symmetries such as $Z_2$.
For instance, the scalar particle may decay to a pair of photons or gluons through  dimension five operators:
\begin{eqnarray}
C_1\frac{S F_{\mu\nu}F^{\mu\nu}}{\Lambda}+C_2 \frac{S G^a_{\mu\nu}G_a^{\mu\nu}}{\Lambda},
\end{eqnarray}
where $C_1\sim C_2 \sim O(1)$ are (unknown) parameters. One should notice that
both operators respect gauge symmetry but break $Z_2$ symmetry.
The life time of the scalar is suppressed by a large cutoff scale ($\Lambda\sim M_{\rm Planck}$)
but certainly much shorter than the age of universe so that we can avoid the strong constraint
from the relic density measurements.

\section{$B\to K SS$ and Invisible Higgs}
In this section We study the constraint
on the interaction term between the Higgs boson and the SM singlet from $B$ decays. Specifically we will
look at $B \to K S S$ decay which currently has the most stringent
experimental upper bound $14 \times 10^{-6}$ \cite{:2007zk}.

The effective Hamiltonian for this decay can be expressed as
\begin{equation}
H_{\rm eff}=\frac{\lambda V_{tb}^* V_{ts}}{2m_h^2} C_s \bar{s}
(1+\gamma_5) b S S~.
\end{equation}
Intuitively, $b \to sSS$ decay can be divided into two processes: first
b quark decays to s quark plus a
off-shell Higgs boson $h$, and subsequently $h$ decays to two light singlets.
From the interaction Lagrangian term $\lambda H^+ H S^2 /2v_0$, with
$H^+ = (\phi^-, (v_0+h-i \phi^0)/\sqrt{2})$, it is easy to show that the Higgs boson
decay $h \to SS$ can proceed through a trilinear term $\lambda h S S /2$. But
as we will see later, another term $\lambda \phi^+ \phi^- S^2/2v_0$ is also crucial
to guarantee the gauge independence of the decay amplitude.

To evaluate the decay amplitude, the Wilson coefficient $C_s$ at
scale $\mu_b={\cal O}(m_b)$ should be known, which can be obtained
by matching the full theory to the effective theory at scale around
$m_W$ to obtain $C_s(m_W)$ and then evolving down to $\mu_b$.
As the above operator does not mix with other effective operators,
the QCD running effects can be obtained straightforwardly with the calculation
of the anomalous dimension of $\bar{s}
(1+\gamma_5) b$ \cite{Dai:1996vg}:
\begin{equation}\label{eq:qcd running}
C_s(\mu_b)=\left (\frac{\alpha_s(\mu_b)}{\alpha_s(m_W)}\right )^{12/23}C_s(m_W)~.
\end{equation}
$C_s(m_W)$ can be obtained
by calculating the diagrams in Fig. 2. Notice that the Higgs boson
does not couple to $s$-quark by taking $m_s=0$.

In Fig. 2, the first eight diagrams represent exactly the intuitive picture that first $b \to s h$,
and then $h \to SS$. Since the later one is a trivial tree level process,
one may first focus on the construction of an one-loop effective $bsh$ vertex
\begin{equation}
         {\cal L}_{bsh}=C_{bsh} V_{tb}^* V_{ts} \bar{s}(1+\gamma_5)b h
\end{equation}
with the coefficient in t'Hooft-Feynman gauge as \cite{Willey:1982mc,Grzadkowski:1983yp}
\begin{equation}
       C_{bsh}(m_W)=\frac{g^2}{(4\pi)^2}\frac{m_b x_t}{8v_0}\left (
       3 +x_h \frac{(3-x_t)(1-x_t)+2x_t(2-x_t)\ln x_t}{(1-x_t)^3} \right )~,
\end{equation}
where $x_t\equiv m_t^2/m_W^2$, $x_h \equiv m_h^2/m_W^2$ with the approximation
$m_b^2/(m_W^2,m_t^2,m_h^2) \simeq 0$. Notice that this expression is
gauge-dependent as the Higgs boson is off-shell. Although the calculation itself
is straightforward, the issues about gauge dependence and renormalization scheme
ambiguities are a bit subtle which were finally settle down by several groups a few years later
\cite{Botella:1986gf}.

\begin{figure}[tb]
\begin{center}
\unitlength 1mm
\begin{picture}(140,81)
\put(0,0){\includegraphics{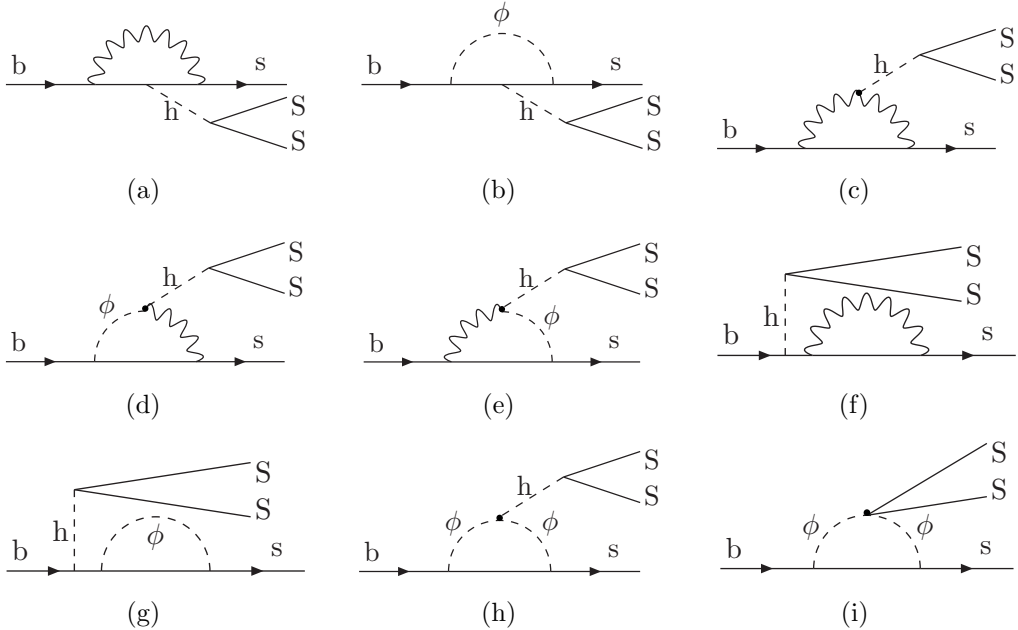}}
\end{picture}
\vspace*{-0.2cm}
\caption{$b$-quark decays to $s$-quark plus two light singlets. The internal quark lines
represent up, charm or top quarks, while the internal dashed lines denote
Higgs boson (h) or unphysical charged goldstone bosons ($\phi$).}
\end{center}
\end{figure}

But for the decay amplitude $b \to s SS$ to be gauge invariant,
the last diagram in Fig. 2, {\it i.e.} Fig. 2(i), has to be included which
(surprisingly at first look) does not contain virtual Higgs boson exchange at all.
Actually Fig. 2(i) arises from the interaction term $\lambda \phi^+ \phi^- S^2/2v_0$.
Therefore strictly speaking, $b \to sSS$ can not be factorized into $b \to sh$ and
$h \to SS$.

Finally, summing all the diagrams, we obtain \footnote{
This expression has been
obtained in \cite{Bird:2004ts}. However, in the derivation, they divided the process $b \to sSS$ into $b\to sh$ and
$h \to SS$. They then evaluate the $bsh$ vertex with the approximation of vanishing
Higgs boson mass. But even with these unrigorousness or approximations, they do obtain finally the correct
expression which is due to the almost completely cancelation between Fig. 2(h) and (i) up
to $O(m_b^2/m_h^2)$. However, generally this kind of cancelation does not happen and
the summation should be of order $m_h^2/m_W^2$, as pointed out by Botella and Lim in
\cite{Botella:1986gf}.}
\begin{equation}
\begin{aligned}
C_s(m_W) &=\frac{g^2}{(4\pi)^2}\frac{3m_b x_t}{8v_0}
\end{aligned}
\end{equation}
The calculation details can be found in the appendix. Here $m_b$ should be evaluated at the scale $m_W$,
but interestingly when combined with the QCD evolution effect of Eq. (\ref{eq:qcd running}), one has\footnote
{We thank the referee for pointing this out to us}
\begin{equation}
   m_b(m_W)\left (\frac{\alpha_s(\mu_b)}{\alpha_s(m_W)}\right )^{12/23}=m_b(m_b)~.
\end{equation}
 Please also note that in \cite{Altmannshofer:2009ma}
authors considered $b \to s S S$ in an effective theory approach, however, with $C_s$ as model independent free parameters.

To get the decay amplitude, the hadronic matrix element
$\langle K^-\vert \bar {s} (1+\gamma_5)b \vert B^- \rangle$ is needed as input,
which can be related to the known form factors through equation of
motion,
\begin{equation}
\begin{aligned}
\langle K^- \vert \bar {s} (1+\gamma_5)b \vert B^- \rangle &=
\frac{q^\mu}{m_b} \langle K^- \vert \bar {s} \gamma_\mu b \vert B^-
\rangle \\
&=\frac{q^\mu}{m_b} \left (
f_+(q^2)(p+l)_\mu+(f_0(q^2)-f_+(q^2))\frac{m_B^2-m_K^2}{q^2}q_\mu
\right )~,
\end{aligned}
\end{equation}
with the light-cone sum rules (LCSR) estimation \cite{Ball:2004ye}
\begin{equation}
\begin{aligned}
f_+(q^2)&=\frac{0.162}{1-q^2/5.41^2}+\frac{0.173}{(1-q^2/5.41^2)^2}\\
f_0(q^2)&=\frac{0.33}{1-q^2/37.46}
\end{aligned}
\end{equation}
As discussed in \cite{Ball:2004ye}, the uncertainty of the $q^2$ dependence of the form factors
have not been fully analyzed in LCSR but likely to be smaller than that at $q^2=0$ which is about $12\%$.
Thus as an rough error estimation we assign a universal $12\%$ uncertainty to the above form factors.

Then, the branching ratio can then be obtained
\begin{equation}
\begin{aligned}
\text{Br}(B \to K SS)&=\frac{\lambda^2 \vert V_{tb}^* V_{ts}
\vert^2}{512 \pi^3 m_B^3 \Gamma_B m_h^4} C_s^2(m_b)  \int
\limits_{4m_S^2}^{(m_B-m_K)^2} \! dq^2 \\
&\hspace*{0.5cm}\langle K^- \vert \bar {s} (1+\gamma_5)b \vert B^-
\rangle^2 \sqrt{q^2-4m_S^2}
\sqrt{\frac{(m_B^2-q^2-m_K^2)^2}{q^2}-4m_K^2}.
\end{aligned}
\end{equation}
Taking as illustration
\begin{equation}
\begin{aligned}
m_h&=130~\mbox{GeV}, \hspace*{1cm} m_S=1~\mbox{GeV},
\end{aligned}
\end{equation}
and with the values \cite{PDG}
\begin{equation}
m_b(m_b)=4.2~\mbox{GeV}, \hspace*{1cm} m_t=171.3~\mbox{GeV}, \hspace*{1cm} A=0.814~, \hspace*{1cm} \lambda_{\rm CKM}=0.2257~
\end{equation}
and $V_{ts}=-A\lambda_{\rm CKM}^2$, we can obtain the branching ratio
\begin{equation}
\text{Br}(B \to K SS)=(0.82 \pm 0.20)\times \left(\frac{\lambda}{1~\rm GeV}\right)^2
\left({130~\text{GeV}\over m_h}\right)^4\times 10^{-10}~,
\end{equation}
where only the form factor uncertainty has been included in the error estimation.

\section{Invisible Higgs}

\begin{figure}[t]
\includegraphics[scale=1.0,width=8cm]{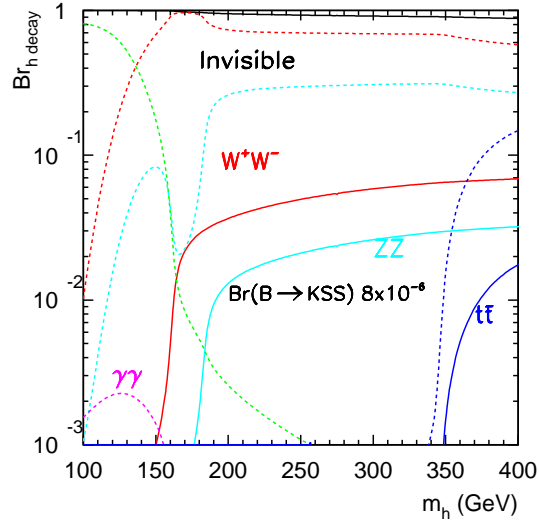}
\caption{Higgs boson decay BR with Invisible decay mode predicted from current upper bound of $B\to K \nu\bar{\nu}$ in solid
lines. \textit{(For comparison, dashed lines are for SM Higgs decay BR.)}}
\label{bound}
\end{figure}

If there exists such light SM singlet scalar, the Higgs decay can be significantly modified.
For $m_S=1$~GeV, we take the upper bound on $\lambda$ derived from the $B\to K\cancel{E}$ as
\begin{eqnarray}
\text{Br}(B\to K SS)&=&0.82\times 10^{-10} \left(\frac{\lambda}{1~\rm GeV}\right)^2
\left({130~\text{GeV}\over m_h}\right)^4  \nonumber \\
&\le& \text{Br}_\text{exp}(B\to K\cancel{E})-\text{Br}_\text{SM}(B\to K\nu\bar{\nu})
\simeq 8 \times 10^{-6}
\end{eqnarray}
and compute the upper bound of partial width for $h\to SS$.
The partial width of Higgs decaying into two scalar is
\beq
\Gamma(h\rightarrow SS) = {\lambda^2\over 32 \pi m_h}\left(1-{4m^2_S\over m^2_h}\right)^{1\over 2},
\eeq
where $\lambda$ is the dimension one coupling and $m_h$, $m_S$ are the Higgs boson mass and hidden
sector scalar mass respectively.
To illustrate the feature,  we scan $m_h$ and plot
in Fig.~\ref{bound} how the Higgs decay BR will be changed due to the
$h\to SS$ decay. The partial width of $h\to SS$
is obtained by taking $m_S=1$ GeV and the $\lambda$ upper
bound value computed for that $m_h$ point.

If $m_h< 150$~GeV, $h\to SS$ completely dominates the Higgs decay and Higgs is only invisible.
Even though the traditional invisible Higgs search can be applied to search for such modes,
it is impossible to identify the resonance
through invisible modes at the LHC.

When $m_h>150$~GeV, the partial width of $h\to SS$ is comparable to the partial widths of conventional channels,
such as $h\to W^+W^-$ or $h\to ZZ$. The multi-lepton searches for Higgs resonance
are still valid but the decay BRs significantly decrease.
If the measured event numbers of $h\to W^+W^-$ or $h\to ZZ$ are below
the expected numbers. There are several possibilities:
\begin{itemize}
\item There are more than one Higgs boson responsible for the $W$ gauge boson mass $M_W$. The vacuum
expectation value for the lightest Higgs boson is much smaller than $v_0$ so that the coupling
$W^+W^-H$ is $gv^\prime$.
\item The production of Higgs boson is suppressed due to new physics. For instance, $gg\to H$ production is less due to the top quark partner in the
triangle loop and significantly cancel the top quark loop.
\item There exists unknown Higgs decay mode which cannot be easily identified. Invisible Higgs mode that we discuss
here falls into this category. Another example is the $h\to \nu N$ decay in some TeV neutrino models \cite{neutrino}.
\end{itemize}

\begin{figure}[t]
\includegraphics[scale=1.0,width=8cm]{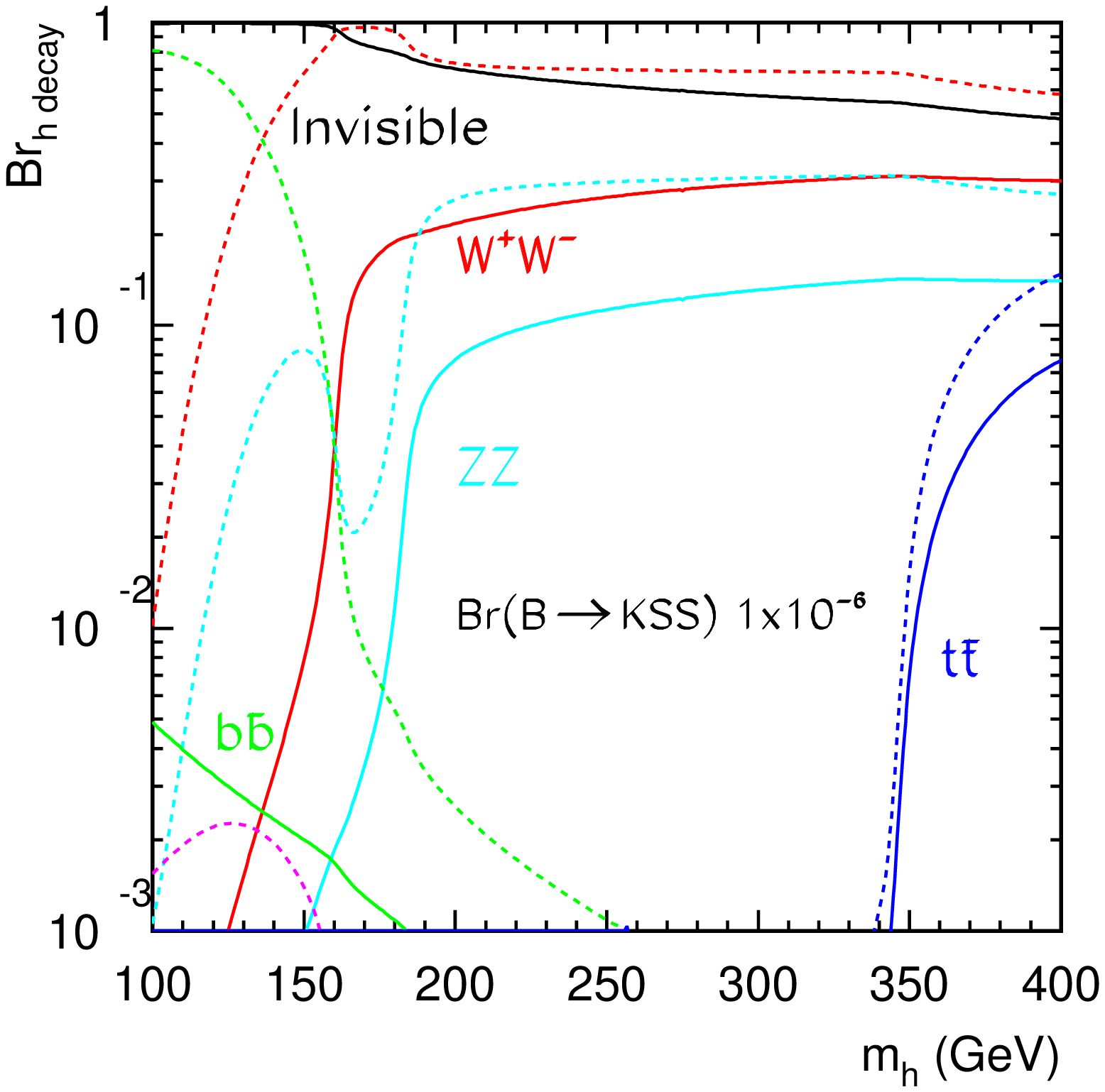}
\includegraphics[scale=1.0,width=8cm]{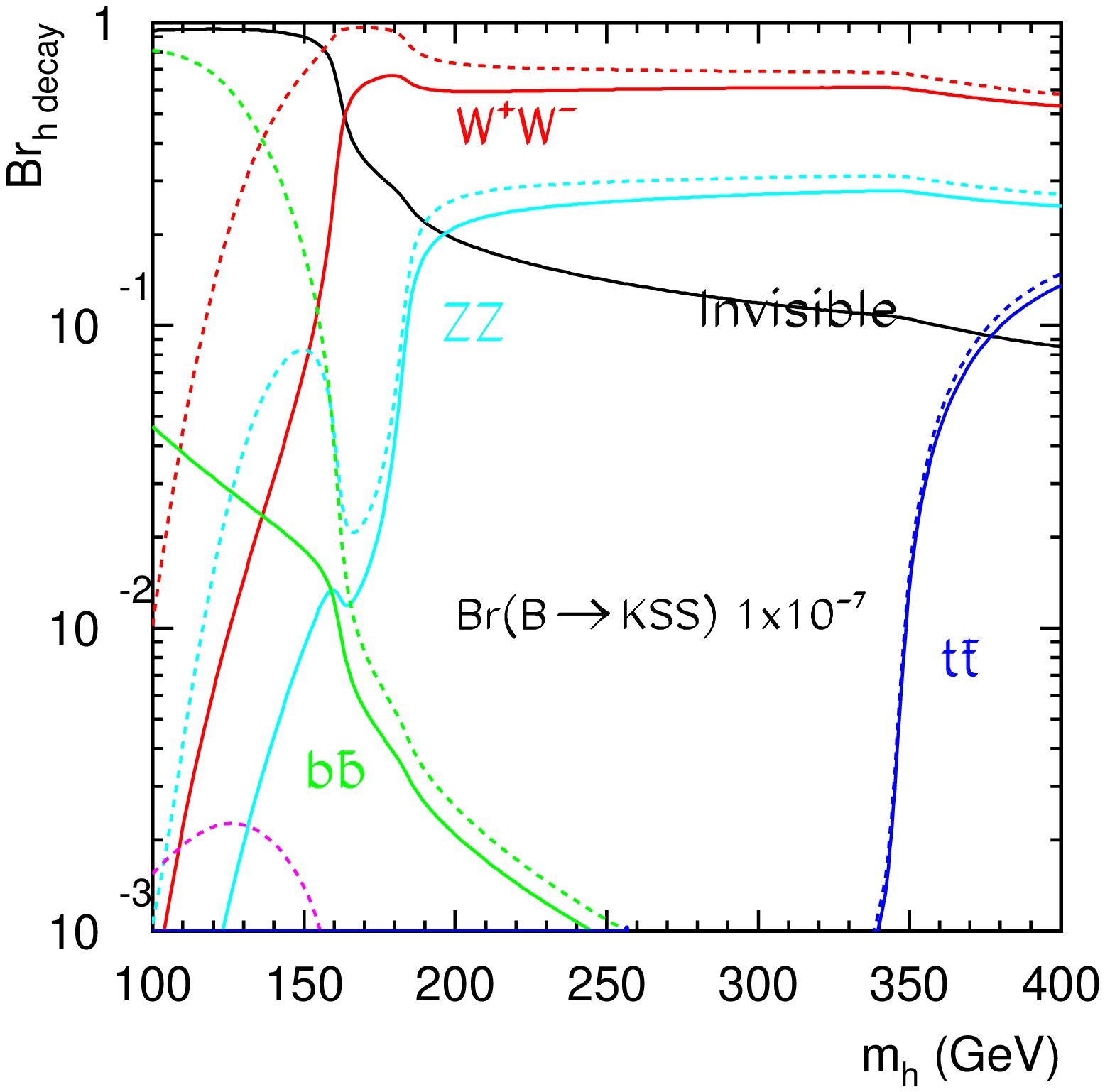}
\caption{Solid lines correspond to the Higgs BR with invisible decay mode predicted from the
upper bound value for $m_S=1$~GeV and Br($B\to K SS) = 1\times 10^{-6}$ or $1\times 10^{-7}$ respectively. Dashed
lines are the standard SM Higgs decay BR. }
\label{future}
\end{figure}

We expect the SuperB or SuperBelle will improve the measurement significantly and
reduce the allowed region of
$\text{Br}(B\to K SS)=\text{Br}_\text{exp}(B\to K\cancel{E})-\text{Br}_\text{SM}(B\to K\nu\bar{\nu})$.
In Fig. \ref{future}, we plot how the Higgs decay BR will change accordingly for $m_S=1$~GeV and
improved bound on $\text{Br}(B\to K SS)$.
As can be seen, if the value of Br($B\to K SS)$ becomes smaller than $2\times 10^{-6}$, it will only change the Higgs decay
before on-shell $WW$ threshold and won't significantly change the
heavy Higgs decay.

\section{Conclusion}

We have studied the contribution of virtual Higgs
in $B\to K \cancel{E}$ by assuming Higgs
coupling to a light SM singlet scalar $S$, $B\to K SS$.
For $M_S= 1$~GeV,
$$
\text{Br}(B \to K SS)=(0.82 \pm 0.20)\times \left(\frac{\lambda}{1~\rm
GeV}\right)^2
\left({130~\text{GeV}\over m_h}\right)^4\times 10^{-10}~.
$$
Given the current experimental bound and subtracting the known SM
contribution,
$$
\text{Br}_\text{exp}(B\to K\cancel{E})-\text{Br}_\text{SM}(B\to K\nu
\bar{\nu})
\simeq 8 \times 10^{-6}~,
$$
we obtain an upper bound on the coupling between the Higgs and
singlet scalar $S$. We take the upper bound value of this coupling
and compute the $h\to SS$ decay partial width. It is interesting that
the partial width of $h\to SS$ decay is at comparable level
when the Higgs mass is close to the $WW$ threshold. Consequently,
Higgs may still be discovered via the conventional Higgs search
channels but with a smaller event number. This will have
some impact on precision measurement of Higgs property.
We expect that the SuperB or SuperBelle experiments
can improve the $B\to K\cancel{E}$ measurement and put a
stringent bound on possible invisible Higgs decay.

We have also studied the possible implication in cosmology from
this scalar. It turns out that for the interesting region of couplings
between
$h$ and $S$, such light scalar may not have
enough annihilation cross section and will then over close
the universe if it is a stable particle. We propose a scenario
where $S$ is not stable in the cosmological scale but only
a stable particle in $B$ decay or collider environments.

\section*{Acknowledgement}

CSK was supported in part by Basic Science Research Program through the NRF of Korea
funded by MOEST (2009-0088395),  in part by KOSEF through the Joint Research Program (F01-2009-000-10031-0),
and in part by WPI Initiative, MEXT, Japan. SC was supported by the World Premier International Research Center
Initiative (WPI initiative) by MEXT and also supported by the Grant-in-Aid for scientific research
(Young Scientists (B) 21740172) from JSPS, Japan.
KW is supported by the World Premier International Research Center Initiative (WPI Initiative), MEXT, Japan.
GZ is supported in part by the
National Science Foundation of China under grant No. 10705024 and
No. 10425525, and by the Scientific Research Foundation for the Returned Overseas
Chinese Scholars, State Education Ministry.

\appendix
\section{Calculation details on $b \to s SS$}
In the calculations, we use the t'Hooft-Feynman gauge $\xi=1$.
$p$, $l$ and $q$ denote the momentum of $b$-quark, $s$-quark and virtual
Higgs boson, respectively. We have taken the approximation $q^2-m_h^2 \simeq -m_h^2$ and
dropped a common factor $\lambda/m_h^2$ in the following expressions. We get
\begin{equation}
\begin{aligned}
{\rm Fig.2(a)}&=
\frac{-ig^2V_{tb}^*V_{ts}}{(4\pi)^2}\bar{s}(l)(1+\gamma_5)b(p)
\frac{m_b}{4v_0}\frac{x_t(x_t^2-1-2x_t\ln x_t)}{(x_t-1)^3}
\end{aligned}
\end{equation}
with $x_t \equiv m_t^2/m_W^2$.
\begin{equation}
\begin{aligned}
{\rm Fig.2(b)}&=
\frac{-ig^2V_{tb}^*V_{ts}}{(4\pi)^2}\bar{s}(l)(1+\gamma_5)b(p)\frac{x_t
m_b}{4v_0}\left
(\frac{1}{\epsilon}-\gamma+\ln 4\pi-\frac{1}{2} \right . \\
& \left. \hspace*{4cm} -2 \int \limits_{x+y \le 1} \! dx  dy
\left(\ln \frac{\Delta_1(x,y)}{\mu^2} +\frac{m_t^2
y}{\Delta_1(x,y)}\right )\right )~
\end{aligned}
\end{equation}
with
\begin{equation}
\begin{aligned}
\Delta_1(x,y)
&\simeq(1-x-y) m_W^2 +(x+y) m_t^2~\nonumber~.
\end{aligned}
\end{equation}
The divergence of Fig. 2(b) can be canceled by that of Fig. 2(g):
\begin{equation}
\begin{aligned}
{\rm Fig.2(g)}
&=\frac{ig^2V_{tb}^*V_{ts}}{(4\pi)^2}\bar{s}(l)(1+\gamma_5)b(p)\frac{x_t
m_b}{4v_0}\\
& \hspace*{3cm}\left (\frac{1}{\epsilon}-\gamma+\ln 4\pi-\int_0^1 \!
dx \ln \frac{x m_t^2+(1-x)m_W^2}{\mu^2} \right )~.
\end{aligned}
\end{equation}
The sum of Figs. 2(b) and (g) then gives (taking the scale $\mu=m_W$)
\begin{equation}
\begin{aligned}
{\rm Fig.2(b+g)}&=\frac{ig^2V_{tb}^*V_{ts}}{(4\pi)^2}\bar{s}(l)(1+\gamma_5)b(p)\frac{x_t^2
m_b}{4v_0} \frac{(3x_t-5)(x_t-1)-2(x_t-2)\ln x_t}{(x_t-1)^3}.
\end{aligned}
\end{equation}
It is clear that for Figs. 2(a),(b),(g), the internal up and charm
quarks contributions are suppressed at least by $m_{u,c}^2/m_t^2$
compared to the virtual top quark contribution and can be safely
neglected.

For Fig. 2(c), the internal top quark contribution is
\begin{equation}
\begin{aligned}
{\rm Fig.2(c)}_t
&=\frac{-ig^2V_{tb}^*V_{ts}}{(4\pi)^2}\bar{s}(l)(1+\gamma_5)b(p)
\frac{m_b}{4v_0}\frac{2x_t^2\ln x_t-(3x_t-1)(x_t-1)}{(x_t-1)^3}~.
\end{aligned}
\end{equation}
But here the internal up and charm quarks
contributions are not suppressed, which can be obtained from the
above expression by taking the limit $x_t \to 0$ and changing the
corresponding CKM factors. We then obtain using the CKM unitarity
condition,
\begin{equation}
\begin{aligned}
{\rm Fig.2(c)}&=\frac{-ig^2V_{tb}^*V_{ts}}{(4\pi)^2}\bar{s}(l)(1+\gamma_5)b(p)
\frac{m_b}{4v_0}\left (\frac{2x_t^2\ln
x_t-(3x_t-1)(x_t-1)}{(x_t-1)^3}-1\right )~.
\end{aligned}
\end{equation}

The virtual top quark contribution to Fig. 2(d) is
\begin{equation}
\begin{aligned}
{\rm Fig.2(d)}_t
&=\frac{-ig^3V_{tb}^*V_{ts}}{(4\pi)^2}\bar{s}(l)(1+\gamma_5)b(p)\frac{m_b}{8m_W}\\
&  \left ( \frac{1}{\epsilon}-\gamma+\ln 4\pi -\frac{1}{2}- \int
  \limits_{x+y \le 1}\! dxdy \left ( 2\ln
  \frac{\Delta_2(x,y)}{\mu^2}+
  \frac{(1+x+y)m_t^2}{\Delta_2(x,y)} \right ) \right )
\end{aligned}
\end{equation}
with
\begin{equation}
\begin{aligned}
\Delta_2(x,y)
             &\simeq x m_t^2+(1-x)m_W^2~.
\end{aligned}
\end{equation}
The divergence here can be canceled when the contributions from the
internal up and charm quarks are included, then we get
\begin{equation}
\begin{aligned}
{\rm Fig.2(d)}&=\frac{ig^3V_{tb}^*V_{ts}}{(4\pi)^2}\bar{s}(l)(1+\gamma_5)b(p)
\frac{x_t m_b}{32m_W}\frac{2x_t(5x_t-6)\ln
x_t-(9x_t-11)(x_t-1)}{(x_t-1)^3}~.
\end{aligned}
\end{equation}

For Fig. 2(e), we have
\begin{equation}
\begin{aligned}
{\rm Fig.2(e)}
&=\frac{-ig^3V_{tb}^*V_{ts}}{(4\pi)^2}\bar{s}(l)(1+\gamma_5)b(p)\frac{x_t
m_b}{32m_W}\frac{2x_t(3x_t-2)\ln x_t-(7x_t-5)(x_t-1)}{(x_t-1)^3}~.
\end{aligned}
\end{equation}
Here the internal up and charm quarks contributions are
again negligibly small due to the ${\cal O}(m_{u,c}^2/m_t^2)$
suppression.
It is easy to show that the contribution of Fig. 2(f) vanishes
by using the equation of motion $\bar{s}(l) \lslash =0$. The cancelation between
Fig. 2(h) and Fig. 2(i) is obvious by approximating the Higgs boson propagator
$i/(q^2-m_h^2)\simeq -i/m_h^2$.

\end{document}